\documentclass{article}
\usepackage{here,graphicx,subfigure}

\begin{document}

\title{A Compact Representation of the Three-Gluon Vertex}

\author{
N. Ahmadiniaz, C. Schubert\\
\\
\it{ Dipartimento di Fisica, Universit\`a di Bologna and INFN},\\
\it{Sezione di Bologna Via Irnerio 46, I-40126 Bologna, Italy}\\
\\
\it {Instituto de F\'{\i}sica y Matem\'aticas}
\\
\it {Universidad Michoacana de San Nicol\'as de Hidalgo}\\
\it {Apdo. Postal 2-82}\\
\it {C.P. 58040, Morelia, Michoac\'an, M\'exico}
\\
\\
Talk given by N. Ahmadiniaz at the\\ 
\it{3rd Young Researchers Workshop}\\
\it{ "Physics Challenges in the LHC Era"}\\
\it{Frascati, May 7 and 10, 2012 }\\
}
\maketitle
\baselineskip=11.6pt
\begin{abstract}
The three-gluon vertex is a basic object of interest in nonabelian gauge theory.
It contains important structural information, in particular on infrared divergences,
and also figures prominently in the Schwinger-Dyson equations. At the one-loop level, it has
been calculated and analyzed by a number of authors. Here we use the worldline formalism
to unify the calculations of the scalar, spinor and gluon loop contributions to the one-loop vertex,
leading to an extremely compact representation. The SUSY - related sum rule found by Binger and
Brodsky follows from an off-shell extension of the Bern-Kosower replacement rules. 
We explain the relation of the structure
of our representation to the low-energy effective action.
 \end{abstract}
\newpage
\section{Introduction}
\label{intro}
The off-shell three-gluon vertex has been under investigation for more than three decades. 
By an analysis of the nonabelian gauge Ward identities, Ball and Chiu\cite{bc} in 1980 found a
form factor decomposition of this vertex which is valid at any order in perturbation theory, with the only restriction that
a covariant gauge be used. At the one-loop level, they also calculated the vertex explicitly for the case of a gluon
loop in Feynman gauge. 
Later Cornwall and Papavassiliou\cite{cp} applied the pinch technique to the 
non-perturbative study of this vertex. 
Davydychev, Osland and Sax \cite{daos} calculated the massive quark contribution 
of the one loop three-gluon vertex.
Binger and Brodsky\cite{bibr} 
calculated the one-loop vertex in the pinch technique 
and found the following SUSY-related identity between its scalar, spinor and gluon loop
contributions,

\begin{equation}
3 \Gamma_{\rm scalar} + 2 \Gamma_{\rm spinor} + \Gamma_{\rm gluon}  = 0\nonumber\\
\label{gammaBB}
\end{equation}

In this talk, I present a recalculation of the scalar, spinor and gluon loop contributions to the
three-gluon vertex using the worldline formalism \cite {bernkos,strassler,rsc,cs} . The vertex is shown in fig. 1 (for the fermion loop case).

\begin{figure}[h]
\hspace{100pt}{\centering
\includegraphics{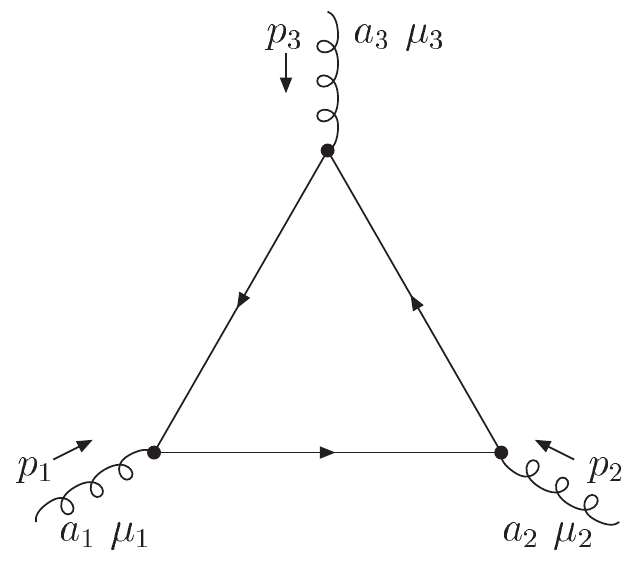}
}

\caption{Three-gluon vertex.}
\label{fig1}
\end{figure}

\noindent
Following the notation of \cite{daos}, we write

\begin{equation}
\Gamma_{\mu_1\mu_2\mu_3}^{a_1a_2a_3}(p_1,p_2,p_3) = -igf^{a_1a_2a_3}\Gamma_{\mu_1\mu_2\mu_3}(p_1,p_2,p_3)
\label{stripcol}
\end{equation}
The gluon momenta are ingoing, such that $p_1+p_2+p_3=0$. 
There are actually two diagrams differing by the two inequivalent orderings of the three gluons along the loop. 
Those diagrams add to produce a factor of two.

\noindent
The Ball-Chiu decomposition of the vertex can be written as
\newpage
\begin{eqnarray}
\Gamma_{\mu_{1}\mu_{2}\mu_{3}}(p_{1},p_{2},p_{3})&=&
A(p_1^2,p_2 ^2;p_3^2)g_{\mu_{1}\mu_{2}}(p_1-p_2)_{\mu_{3}}
+B(p_1^2,p_2 ^2;p_3^2)g_{\mu_{1}\mu_{2}}
(p_1+p_2)_{\mu_{3}}\nonumber\\
&&+C(p_1^2,p_2 ^2;p_3^2)\Big\lbrack p{_1}_{\mu_2}p{_2}_{\mu_1}-(p_1\cdot p_2)g_{\mu_{1}\mu_{2}}\Bigl\rbrack(p_1-p_2)_{\mu_3}\nonumber\\
&&+\frac{1}{3}S( p_1^2,p_2 ^2,p_3^2)\Bigl\lbrack p{_1}_{\mu_3}p{_2}_{\mu_1}p{_3}_{\mu_2}+p{_1}_{\mu_2}p{_2}_{\mu_3}p{_3}_{\mu_1}\Bigl\rbrack\nonumber\\
&&+F( p_1^2,p_2 ^2;p_3^2)\Bigl\lbrack (p_1\cdot p_2)g_{\mu_1\mu_2}-{p_1}_{\mu_2}p{_2}_{\mu_1}\Bigr\rbrack
\Bigl\lbrack p{_1}_{\mu_3}(p_2\cdot p_3)-p{_2}_{\mu_3}(p_1\cdot p_3)\Bigl\rbrack
\nonumber\\
&&+H(p_1^2,p_2 ^2,p_3^2)
\Big\{-g_{\mu_1\mu_2}\Bigl\lbrack p{_1}_{\mu_3}(p_2\cdot p_3)-p{_2}_{\mu_3}(p_1\cdot p_3)\Bigl\rbrack\nonumber\\
&&\qquad\qquad\qquad\quad +\frac{1}{3}\Big\lbrack p{_1}_{\mu_3}p{_2}_{\mu_1}p{_3}_{\mu_2}-p{_1}_{\mu_2}p{_2}_{\mu_3}p{_3}_{\mu_1}\Bigl\rbrack\Big\}\nonumber\\
&&+\Big\{ {\mbox {cyclic permutations of } (p_1,\mu_1),(p_2,\mu_2),(p_3,\mu_3)}\Big\}\nonumber\\
\label{B.C}
\end{eqnarray}
Here  the $A$, $C$ and $F$ functions are symmetric in the first two arguments, $B$ antisymmetric, and H(S) are totally (anti)symmetric with respect to interchange of any pair of arguments. Note that the $F$ and $H$ functions are totally transverse, i.e., they vanish when contracted with any of $p_{1\mu_1}$, $p_{2\mu_2}$ or $p_{3\mu_3}$.\\

\section{The scalar loop case}
\label{scalar}

In the worldline formalism the three-gluon amplitude for the scalar loop case is represented as  \cite{strassler,cs} 
\begin{eqnarray}
&&\Gamma^{a_{1}a_{2}a_{3}}_{\rm scalar}[p_{1},\varepsilon_{1};p_{2},\varepsilon_{2};p_{3},\varepsilon_{3}]
 =(-ig)^{3}\mbox{tr}(T^{a_{1}}T^{a_{2}}T^{a_{3}})\int^{\infty}_0 \frac{dT}{T}{\rm{e}}^{-m^2 T}\nonumber\\
 &\times&\int\mathcal{D} x(\tau) \int_0 ^Td\tau_{1}\varepsilon_{1}\cdot\dot{x}_{1}\,{\rm{e}}^{ip_{1}\cdot x_{1}}\int_0^{\tau_{1}} d\tau_{2}
 \varepsilon_{2}\cdot\dot{x}_{2}\,{\rm{e}}^{ip_{2}\cdot x_{2}}\varepsilon_{3}\cdot\dot{x}_{3}\,
 {\rm{e}}^{ip_{3}\cdot x_{3}}~{\rm{e}}^{-\int_0^T \frac{\dot{x}^{2}}{4}}~~~\nonumber\\
\label{bk}
 \end{eqnarray}
 Here  
 $T$  is the total proper time of the loop particle, $m$ the mass of the loop particle, 
 $T^a$ a generator of the gauge group in the representation of the scalar, and $\int\mathcal{D}(x)$ an integral over closed trajectories 
 in Minkowski space-time with periodicity $T$.
 Although our calculation will be off-shell, we introduce gluon polarization vectors $\varepsilon_{i}$ as a book-keeping device. 
 Each gluon is represented by a vertex operator  $\int d\tau\varepsilon\cdot\dot{x}\,{\rm{e}}^{ip\cdot x}\rm$.
 Translation inveriance in proper-time has been used to set $\tau_3=0$.
 
 \noindent
The path integral (\ref{bk}) is Gaussian so that its evaluation requires only the 
standard combinatorics of Wick contractions and the appropriate 
Green's function,

\vspace{-7pt}

 \begin{eqnarray}
 <x^{\mu_{1}}(\tau_{1})x^{\mu_{2}}(\tau_{2})> = \! - G_{B12}g^{\mu_{1}\mu_{2}},\,
G_{B12}:=G_{B}(\tau_{1},\tau_{2})=|\tau_{1}-\tau_{2}|-\frac{(\tau_{1}-\tau_{2})^2}{T}~~~~~~~~~~~~~\nonumber\\
 \label{G}
 \end{eqnarray} 
 \vspace{-15pt}
 
In this formalism structural simplification can be expected from the removal of all 
second derivatives $\ddot{G}_{Bij}$'s 
, appearing after 
the Wick contractions, by suitable integrations by part (IBP). After doing this we have (see\cite{cs} for the 
combinatorial details of the Wick contraction and IBP procedure)
\begin{eqnarray}
\Gamma_{\rm scalar}&=&\frac{g^3}{(4\pi)^{D/2}}(\Gamma_{\rm scalar}^3+\Gamma_{\rm scalar}^2+\Gamma_{\rm scalar}^{\rm bt})\nonumber\\
\Gamma_{\rm scalar}^3 &=&- {\mbox tr}(T^{a_{1}}[T^{a_{2}},T^{a_{3}}])\int_{0}^{\infty} \frac{dT}{T^{\frac{D}{2}}}{\rm e}^{-m^2 T}\int_{0}^{T}d\tau_{1}\int_0^{\tau_{1}}d\tau_{2}\, Q_3^3\vert_{\tau_3=0}~ 
{\rm e}^{(G_{B12}p_{1}\cdot p_{2}+G_{B13}p_{1}\cdot p_{3}+G_{B23}p_{2}\cdot p_{3})}\nonumber\\
\Gamma_{\rm scalar}^2 &=& \Gamma_{\rm scalar}^3(Q_3^3\to Q_3^2) \nonumber\\
\Gamma_{\rm scalar}^{\rm bt} &=& - {\mbox tr}(T^{a_{1}}[T^{a_{2}},T^{a_{3}}])
\int_{0}^{\infty} \frac{dT}{T^{\frac{D}{2}}}{\rm e}^{-m^2 T}\int_{0}^{T}d\tau_{1}
\dot{G}_{B12}\dot{G}_{B21} 
\Bigl\lbrack\varepsilon_3\cdot f_1\cdot\varepsilon_2~
{\rm e}^{G_{B12}p_{1}\cdot (p_{2}+p_{3})} 
+{\rm 2\,perm}
\Bigr\rbrack\nonumber\\
Q_{3}^3&=&\dot{G}_{B12}\dot{G}_{B23}\dot{G}_{B31}{\mbox tr}(f_1f_2f_3) \nonumber\\
Q_{3}^2&=&\frac{1}{2}\dot{G}_{B12}\dot{G}_{B21}{\mbox tr} (f_1f_2)
\sum_{k=1,2}\dot{G}_{B3k}\varepsilon_{3}\cdot p_{k}+{\rm 2\,perm}\nonumber\\
\label{Q3}
\end{eqnarray}
\vspace{-10pt}

\noindent
The abelian field strength tensors $f^{\mu\nu}_i:=p_{i}^{\mu}\varepsilon_{i}^{\nu}-\varepsilon_{i}^{\mu}p_{i}^{\nu}$ 
appear automatically in the IBP procedure. 
The $\Gamma_{\rm scalar}^{\rm bt}$'s are boundary terms of the IBP.

\noindent
We rescale to the unit circle, $\tau_{i}=Tu_{i}, i=1,2,3,$ and rewrite these integrals in term of the standard \emph{Feynman/Schwinger} parameters, related 
to the $u_i$ by
\begin{eqnarray}
u_{1}=\alpha_{2}+\alpha_{3}~~~,~~~u_{2}=\alpha_{3}~~~,~~~
u_{3}=0 ~~~,~~~\alpha_{1}+\alpha_{2}+\alpha_{3}=1
\label{alpha}
\end{eqnarray}
For the scalar case, we find

\begin{eqnarray}
\Gamma_{\rm scalar} &=& \frac{g^3}{(4\pi)^{\frac{D}{2}}}\mbox{tr}(T^{a_{1}}[T^{a_{2}},T^{a_{3}}])(\gamma_{\rm scalar}^3 + \gamma_{\rm scalar}^2 + \gamma_{\rm scalar}^{\rm bt})\nonumber\\
\gamma_{\rm scalar}^3 &=& \Gamma\bigl(3-\frac{D}{2}\bigr)\mbox{tr} (f_1f_2f_3)
I^D_{3,B}(p_1^2,p_2^2,p_3^2)
\nonumber\\
\gamma_{\rm scalar}^2 &=& \frac{1}{2}\Gamma\bigl(3-\frac{D}{2}\bigr)\Bigl\lbrack
\mbox{tr} (f_1f_2)\Bigl(\varepsilon_3\cdot p_1 I^D_{2,B}(p_1^2,p_2^2,p_3^2)-\varepsilon_3\cdot p_2 I^D_{2,B}(p_2^2,p_1^2,p_3^2)\Bigr)+{\rm 2\,perm}\Bigr\rbrack\nonumber\\
\gamma_{\rm scalar}^{\rm bt} &=& -\Gamma\bigl(2-\frac{D}{2}\bigr)
\Bigl\lbrack\varepsilon_3 \cdot f_1\cdot\varepsilon_2 I^D_{{\rm bt},B}(p_1^2)+{\rm 2\,perm}
\Bigr\rbrack\nonumber\\
\label{gammas0fin}
\end{eqnarray}
\newpage
where
\begin{eqnarray}
I_{3,B}^D(p_1^2,p_2^2,p_3^2) &=& \int_0^1d{\alpha}_1d{\alpha}_2d{\alpha}_3\delta(1-{\alpha}_1-{\alpha}_2-{\alpha}_3)\nonumber\\
&\times&\frac{(1-2{\alpha}_1)(1-2{\alpha}_2)(1-2{\alpha}_3)}{\Bigl( m^2 + {\alpha}_1{\alpha}_2p_1^2+{\alpha}_2{\alpha}_3p_2^2+{\alpha}_1{\alpha}_3p_3^2\Bigr)^{3-\frac{D}{2}}}
\nonumber\\
I_{2,B}^D(p_1^2,p_2^2,p_3^2) &=& \int_0^1d{\alpha}_1d{\alpha}_2d{\alpha}_3\delta(1-{\alpha}_1-{\alpha}_2-{\alpha}_3)\nonumber\\
&\times&\frac{(1-2{\alpha}_2)^2(1-2{\alpha}_1)}{\Bigl( m^2 + {\alpha}_1{\alpha}_2p_1^2+{\alpha}_2{\alpha}_3p_2^2+{\alpha}_1{\alpha}_3p_3^2\Bigr)^{3-\frac{D}{2}}}\nonumber\\
I_{bt,B}^D(p^2) &=& \int_0^1d{\alpha}\frac{(1-2{\alpha})^2}{\bigl( m^2 + {\alpha}(1-{\alpha})p^2\bigr)^{2-\frac{D}{2}}}\nonumber\\
\label{Ibt}
\end{eqnarray}
\section{Fermion and gluon loop calculations}
\label{wlpir}
By an off-shell generalization of the Bern-Koswer replacement rules \cite{bernkos}, 
whose correctness for the case at hand we have verified,
one can get the results for the spinor and gluon loop from the scalar loop one 
simply by replacing

\begin{eqnarray}
&&\Gamma_{\rm scalar}\rightarrow\Gamma_{\rm spinor}: ~~~~ I_{\lbrace3,2,{\rm bt}\rbrace,B}^D\rightarrow I_{\lbrace 3,2,{\rm bt}\rbrace,B}^D
-I_{\lbrace 3,2,{\rm bt}\rbrace,F}^D\nonumber\\
&&\Gamma_{\rm scalar}\rightarrow\Gamma_{\rm gluon}: ~~~~~~I^D_{\lbrace 3,2,{\rm bt}\rbrace,B}\rightarrow I^D_{\lbrace 3,2,{\rm bt}\rbrace,B}
-4I^D_{\lbrace 3,2,{\rm bt}\rbrace,F}\nonumber\\
\label{gamma}
\end{eqnarray}
where the $I^D _{(\cdot)F}$'s are three integrals similar to the $I^D _{(\cdot)B}$'s above (for the
spinor loop one must also multiply by a global factor of $-2$).

\noindent
From (\ref{gamma}) we immediately recover the Binger-Brodsky identity eq.(\ref{gammaBB}).

\section{Comparison with the effective action} 
Finally let us compare our results with the low energy expansion of the QCD effective action induced by a scalar loop,
\begin{equation}
\Gamma_{{\rm scalar}}[ F]=\int_{0}^\infty\frac{dT}{T}\frac{{{\rm e}}^{-m^2 T}}{(4\pi T)^{\frac{D}{2}}}{{\rm tr}}\int dx_{0}\sum_{n=2}^{\infty}\frac{(-T)^n}{n!}{{\rm O}}_{n}[ F]
\end{equation}
For our comparison we need only $O_2$ and $O_3$ which are\cite{cs}
\begin{eqnarray}
O_2=-\frac{1}{6}F_{\kappa\lambda}F^{\kappa\lambda}~~~,~~~
O_3=-\frac{2}{15}i F_{\kappa}^{~\lambda}F_{\lambda}^{~\mu}F_{\mu}^{~\kappa}-\frac{1}{20}D_{\kappa}F_{\lambda\mu}D^{\kappa}F^{\lambda\mu}~~
\end{eqnarray}
where
\begin{eqnarray}
F_{\mu\nu}=f_{\mu\nu}+ig[A_{\mu},A_{\nu}]~~,~~~
f_{\mu\nu}=\partial_{\mu}A_{\nu}-\partial_{\nu}A_{\mu}~~,~~
D_{\mu}=\partial_{\mu}+igA_{\mu}
\label{fields}
\end{eqnarray}
We can recognize the correspondences

\begin{eqnarray}
&&\gamma_{(\cdot)}^3\leftrightarrow F_{\kappa}^{~\lambda}F_{\lambda}^{~\mu}F_{\mu}^{~\kappa}=f_{\kappa}^\lambda f_{\lambda}^\mu f_{\mu}^\kappa+{\rm higher~point~terms}\nonumber\\
&&\gamma_{(\cdot)}^2\leftrightarrow (\partial+ig\underbrace {A)F(\partial}+igA)F\nonumber\\
&&\gamma_{(\cdot)}^{\rm bt}\leftrightarrow(f+ig\underbrace{[A,A])(f}+ig[A,A])\nonumber\\
\label{comparison}
\end{eqnarray}

\section{Conclusions and outlook}

In our recalculation of the scalar, spinor and gluon contributions to the one-loop three gluon vertex
we have achieved a significant improvement over previous calculations both in efficiency and
compactness of the result. This improvement is in large part due to the replacement rules (\ref{gamma})
whose validity off-shell we have verified. Details and a comparison with the Ball-Chiu decomposition will
be presented elsewhere.
We believe that along the lines presented here even a first calculation of the four-gluon vertex would be feasible.

\end{document}